\documentclass[prb,showpacs,twocolumn,aps,superscriptaddress,a4paper]{revtex4}
\usepackage{dcolumn}
\usepackage{amsmath}
\usepackage{graphicx}
\usepackage{latexsym}  
\usepackage{amsfonts}  
\usepackage{amssymb}  
\DeclareGraphicsExtensions{.pdf,.gif,.jpg,.eps}  
  
\newcommand{\be}{\begin{equation}}  
\newcommand{\ee}{\end{equation}}
\newcommand{\beq}{\begin{eqnarray}}  
\newcommand{\eeq}{\end{eqnarray}}

\begin{document}  
      
\def\bbe{\mbox{\boldmath $e$}}  
\def\bbf{\mbox{\boldmath $f$}}      
\def\bg{\mbox{\boldmath $g$}}  
\def\bh{\mbox{\boldmath $h$}}  
\def\bj{\mbox{\boldmath $j$}}  
\def\bq{\mbox{\boldmath $q$}}  
\def\bp{\mbox{\boldmath $p$}}  
\def\br{\mbox{\boldmath $r$}}      
  
\def\bone{\mbox{\boldmath $1$}}      
  
\def\dr{{\rm d}}  
  
\def\tb{\bar{t}}  
\def\zb{\bar{z}}  
  
\def\tgb{\bar{\tau}}

\def\bC{\mbox{\boldmath $C$}}  
\def\bG{\mbox{\boldmath $G$}}  
\def\bH{\mbox{\boldmath $H$}}  
\def\bK{\mbox{\boldmath $K$}}  
\def\bM{\mbox{\boldmath $M$}}  
\def\bN{\mbox{\boldmath $N$}}  
\def\bO{\mbox{\boldmath $O$}}  
\def\bQ{\mbox{\boldmath $Q$}}  
\def\bR{\mbox{\boldmath $R$}}  
\def\bS{\mbox{\boldmath $S$}}  
\def\bT{\mbox{\boldmath $T$}}  
\def\bU{\mbox{\boldmath $U$}}  
\def\bV{\mbox{\boldmath $V$}}  
\def\bZ{\mbox{\boldmath $Z$}}  
  
\def\bcalS{\mbox{\boldmath $\mathcal{S}$}}  
\def\bcalG{\mbox{\boldmath $\mathcal{G}$}}  
\def\bcalE{\mbox{\boldmath $\mathcal{E}$}}  
  
\def\bgG{\mbox{\boldmath $\Gamma$}}  
\def\bgL{\mbox{\boldmath $\Lambda$}}  
\def\bgS{\mbox{\boldmath $\Sigma$}}  
  
\def\bgr{\mbox{\boldmath $\rho$}}  
  
\def\a{\alpha}  
\def\b{\beta}  
\def\g{\gamma}  
\def\G{\Gamma}  
\def\d{\delta}  
\def\D{\Delta}  
\def\e{\epsilon}  
\def\ve{\varepsilon}  
\def\z{\zeta}  
\def\h{\eta}  
\def\th{\theta}  
\def\k{\kappa}  
\def\l{\lambda}  
\def\L{\Lambda}  
\def\m{\mu}  
\def\n{\nu}  
\def\x{\xi}  
\def\X{\Xi}  
\def\p{\pi}  
\def\P{\Pi}  
\def\r{\rho}  
\def\s{\sigma}  
\def\S{\Sigma}  
\def\t{\tau}  
\def\f{\phi}  
\def\vf{\varphi}  
\def\F{\Phi}  
\def\c{\chi}  
\def\w{\omega}  
\def\W{\Omega}  
\def\Q{\Psi}  
\def\q{\psi}  
  
\def\ua{\uparrow}  
\def\da{\downarrow}  
\def\de{\partial}  
\def\inf{\infty}  
\def\ra{\rightarrow}  
\def\bra{\langle}  
\def\ket{\rangle}  
\def\grad{\mbox{\boldmath $\nabla$}}  
\def\Tr{{\rm Tr}}  
\def\Re{{\rm Re}}  
\def\Im{{\rm Im}}

\title{The Role of Bound States in Time-Dependent Quantum Transport} 
 
\author{E. Khosravi}  
\affiliation{Institut f\"ur Theoretische Physik, Freie Universit\"at Berlin,   
Arnimallee 14, D-14195 Berlin, Germany}  
\affiliation{European Theoretical Spectroscopy Facility (ETSF)} 
\author{S. Kurth}  
\affiliation{Institut f\"ur Theoretische Physik, Freie Universit\"at Berlin,   
Arnimallee 14, D-14195 Berlin, Germany}  
\affiliation{European Theoretical Spectroscopy Facility (ETSF)}    
\author{G. Stefanucci}   
\affiliation{Department of Physics, University of Rome Tor Vergata, 
Via della Ricerca Scientifica 1, 00133 Rome, Italy}
\affiliation{European Theoretical Spectroscopy Facility (ETSF)}  

\author{E.K.U. Gross}  
\affiliation{Institut f\"ur Theoretische Physik, Freie Universit\"at Berlin,   
Arnimallee 14, D-14195 Berlin, Germany}  
\affiliation{European Theoretical Spectroscopy Facility (ETSF)}

\date{\today}  
  
\begin{abstract}
Charge transport through a nanoscale junction coupled to two macroscopic 
electrodes is investigated for the situation when bound states are present. 
We provide numerical evidence that bound states give rise to persistent, 
non-decaying current oscillations in the junction. We also show that 
the amplitude of these oscillations can exhibit a strong 
dependence on the history of the applied potential as well as on the 
initial equilibrium configuration. Our simulations allow for 
a quantitative investigation of several transient features. We also discuss the existence of different time-scales and address their microscopic origin.

\end{abstract}

  
\maketitle  
  
\section{Introduction}  
\label{intro}

In order to describe electronic transport through mesoscopic or nanoscopic 
devices, a quantum description of transport is essential. A seminal quantum 
theory of transport is the Landauer-B\"uttiker formalism,\cite{Landauer:57,Buettiker:86}
which expresses the conductance of a device 
in terms of the quantum-mechanical transmittance of (non-interacting) 
electrons at the Fermi energy.

In recent years and spurred by experimental 
progress in transport measurements through single 
molecules,\cite{ReedZhouMullerBurginTour:97} the Landauer-B\"uttiker 
formalism has been combined
\cite{Lang:95,ht.1995,SeminarioZacariasTour:98,tgw.2001,pplv.2001,XueDattaRatner:02,BrandbygeMozosOrdejonTaylorStokbro:02,EversWeigendKoentopp:04,flss.2005}
with (static) density functional theory which allows to take the atomistic 
structure of both the molecule and the contacts into account. For a recent 
critical review of this methodology, the reader is referred to 
Ref.~\onlinecite{KoentoppChangBurkeCar:08}. 

The Landauer-B\"uttiker formalism focusses on the description of steady-state 
transport and assumes that for a system which is driven out of equilibrium 
by a dc bias, a dc current will eventually develop, which means that the dynamical formation 
of the steady state is not proved but rather taken for granted. The 
question how the system dynamically reaches a steady state has been 
investigated both 
numerically\cite{KurthStefanucciAlmbladhRubioGross:05,bsdv.2005,sssbht.2006,sbhdv.2007}
and theoretically.\cite{Cini:80,StefanucciAlmbladh:04-2} Using 
non-equilibrium Green functions (NEGF) techniques it has been 
shown\cite{StefanucciAlmbladh:04-2} that the total current  (and 
density) approaches a steady value provided the local 
density of states is smooth in the device region. Such value 
is 1) in agreement with the 
Landauer formula and 2) independent of the initial equilibrium 
configuration and the history of the applied bias.
For a steady state to develop the condition on the local 
density of states excludes the presence of bound 
states. Recently, the inclusion of bound states in time-dependent quantum 
transport has been studied 
in Ref.~\onlinecite{DharSen:06} and further been addressed in subsequent 
work.\cite{Stefanucci:07} There it is demonstrated that if the dc biased 
Hamiltonian supports two or more bound states, the long-time limit of 
the current consists of two terms: a steady-state 
contribution given by the Landauer formula and an additional, 
dynamical contribution responsible for undamped current oscillations.
The frequencies of these oscillations are given by the differences 
between two bound-state energies and, interestingly, the amplitudes 
depend on both the initial state and history of the time-dependent 
perturbation. 

In the present work, the history as well as the initial-state dependence 
of the dynamical part of the current is investigated numerically in detail. 
As a tool for our numerical calculations we use a recently developed algorithm 
\cite{KurthStefanucciAlmbladhRubioGross:05} which allows for the time 
propagation of quantum transport systems according to the Schr\"odinger equation. 

The paper is organized as follows. In Section \ref{pract} we summarize 
the results of Ref.~\onlinecite{Stefanucci:07} which are relevant for 
the discussion of our findings and we briefly describe the central 
ideas of the time-propagation algorithm. In Section \ref{num} we present 
our numerical results which not only confirm the existence of the undamped 
current oscillations but also allow to identify additional internal 
transitions contributing to the transient behavior of the driven system. 
We investigate the dependence of the current oscillations on various 
parameters and initial conditions and provide 
theoretical explanations of the observed behavior. Finally, we 
recapitulate our main results in Section \ref{concl}. 

\section{Two Approaches to Time-Dependent Transport}  
\label{pract}   

In this Section we briefly describe two alternative approaches to 
time-dependent transport in a typical electrode-device-electrode geometry: 
non-equilibrium Green functions (NEGF) and direct solution of the 
time-dependent Schr\"odinger equation.  As was already pointed out 
within the former approach,\cite{DharSen:06,Stefanucci:07}  quantum transport in systems of 
non-interacting electrons exhibits persistent current- (and density-) 
oscillations if two or more bound states are present in the biased system. 
Here, we use the latter approach to address several issues about 
such bound-state 
oscillations. A particularly interesting feature 
of them is the fact that their amplitude depends 
on the entire time evolution as the system is driven out of equilibrium 
(memory effects). 

\subsection{Non-equilibrium Green functions}
\label{thp}

We consider a quantum system of non-interacting electrons which consists 
of a central device (e.g., a quantum point contact or a single molecule plus 
a few atomic layers of the left and right electrodes) and two 
semi-infinite reservoirs (left and right electrodes). As initial 
state we use the one proposed by Cini:\cite{Cini:80} all parts of the system, 
i.e., left lead (region $L$), central device (region $C$) and right lead 
(region $R$), are initially (at $t\leq0$) connected and in a well defined 
equilibrium configuration with a {\em unique} temperature and 
chemical potential (thermodynamic consistency). In this initial state, the 
charge density of the electrodes is perfectly balanced and no current flows 
through the junction. 

For non-interacting electrons at zero temperature, the initial state is a 
Slater determinant of eigenstates 
of the entire contacted system with eigenenergies smaller than the Fermi 
energy. At time $t>0$ the system is driven out of equilibrium by exposing 
it to an external time-dependent potential which is local in time and space. 
For example, we may switch on an electromotive force in such a way 
that the potential drop is entirely limited to the central region.
The boundaries of the open quantum system are chosen in a way that the 
density outside the region $C$ is accurately described by an 
equilibrium bulk density. The time-dependent perturbation may cause a current 
flow through the device. The total current from region $\a=L,R$ can be calculated 
from time derivative of the total number of particles in $\a$:
\begin{equation}
I_{\a}(t)= - e \int_{\a}\dr \br\;\frac{{\rm d}}{{\rm d}t}n(\br,t),
\quad\a=L,R,
\label{cudft}
\end{equation}
where $n(\br,t)$ is the time-dependent electron density and the space integral 
extends over region $\a$ ($e$ is the electron charge). 
Assuming no direct coupling between the left and right 
electrodes, the single-particle Hamiltonian of the entire, contacted system 
can be written as:
\begin{equation}  
    \bH(t)=\left[  
    \begin{array}{ccc}  
	\bH_{LL}(t) & \bH_{LC} & 0 \\  
	\bH_{CL} & \bH_{CC}(t) & \bH_{CR} \\  
	0 & \bH_{RC} & \bH_{RR}(t)  
    \end{array}  
    \right]  .
    \label{ham}  
\end{equation}  
The diagonal blocks of the above matrix are obtained by projecting the full 
Hamiltonian $\bH$ onto the corresponding region. The off-diagonal 
blocks in Eq.~(\ref{ham}) account for the coupling between the device region 
$C$ and the leads and, for simplicity, we assume them to be 
time-independent. For instance, in a real-space representation using a 
finite-difference discretization of the kinetic energy, the off-diagonal 
elements of $\bH$ are simply given by the off-diagonal elements of the 
kinetic energy operator. (Model systems with time-dependent couplings were 
studied, e.g., in Ref.~\onlinecite{MoldoveanuGudmundssonManolescu:07}.)

One way to deal with non-equilibrium problems is provided by the NEGF theory. 
 From the equation of motion of the Keldysh-Green function one can 
rewrite the current $I_{\a}(t)$ of Eq.~(\ref{cudft})
in terms of the lesser Green function projected onto different 
subregions as:
\begin{equation}
 I_{\a}(t)=2e {\rm Re}\,{\rm Tr} [\mathbf{G}_{C \a}^{<}(t,t) 
 \bH_{\a C}],
 \end{equation}
where ${\rm Tr}$ denotes the trace over a complete set of states in the 
central region. The lesser Green function can be 
expressed \cite{bnh.1976,Cini:80,d.1984,StefanucciAlmbladh:04,StefanucciAlmbladh:04-2} in terms of 
retarded and advanced Green functions as 
\begin{equation}
 \mathbf{G}^{<}(t;t')=\mathbf{G}^{R}(t;0) \mathbf{G}^{<}(0;0) 
\mathbf{G}^{A}(0;t').
\end{equation}
The initial condition is $\mathbf{G}^{<}(0;0)=  i f(\bH^{0})$ where  
$f(\w)=(e^{\b(\w-\m)}+1)^{-1}$ is the Fermi distribution function and $\bH^0$ 
is the (time-independent) Hamiltonian for $t<0$. 

It can be shown \cite{StefanucciAlmbladh:04} that in a dc biased system 
the total time-dependent current approaches 
a steady value {\em provided the local density of states in 
region $C$ is smooth}. In this case, the steady current 
is given by:
 \begin{equation}
I_{L}^{(S)}= \lim_{t \to \infty} I_{\a}(t) = e \int \frac {d \w} {2\p} 
[f(\w -U _{L}^{\inf})-f(\w -U _{R}^{\inf})]T(\w).
\label{stcurr} 
\end{equation}
In the above equation $U_{\a}^{\inf}$ is the value approached by the bias 
in lead $\a$ when $t \to \inf$ and 
$T(\w)=\Tr[\mathbf{G}^{R}_{CC}(\w) \mathbf{\G}_{L}(\w) 
\mathbf{G}^{A}_{CC}(\w) \mathbf{\G}_{R}(\w)]$, where $\mathbf{\G}_{\a}(\w)= 
-2 {\rm Im}[\bgS_{\a}^{R}(\w)]$ and $\mathbf{G}^{R/A}_{CC}$ are the retarded 
and advanced Green functions projected in region $C$. 
$\bgS_{\a}^{R}(\w)=\bH_{C\a} \bg^R_{\a \a}(\w) \bH_{\a C} $ is the embedding 
self energy with the retarded Green function of 
lead $\a$, $\bg^R_{\a \a}(\w) = \left( \w - \bH_{\a \a}^0 
-U^{\inf}_{\a}+ i 0^+ \right)^{-1}$ .
The steady current does not depend on the initial Hamiltonian (the memory 
of different initial conditions is completely washed out) and is also 
independent of the history of the applied bias (memory-loss theorem).\cite{StefanucciAlmbladh:04-2}
 
The above scenario changes drastically if the Hamiltonian 
$\bH^{\inf}:= \lim_{t \to \inf}\bH(t)$ has two or more bound eigenstates. 
In this case the long-time limit of the current has two contributions:\cite{Stefanucci:07}
\begin{equation}
\lim_{t\rightarrow \inf} 
I_{\alpha}(t)=I_{\alpha}^{(S)}+I_{\alpha}^{(D)}(t) \; .
\label{s+dcurr} 
\end{equation} 
In addition to the steady-state contribution $I_{\a}^{(S)}$ given by 
Eq.~(\ref{stcurr}) one finds  
a dynamical, explicitly time-dependent contribution 
$I_{\alpha}^{(D)}$ which can be written as  
\begin{equation}
I_{\a} ^{(D)}(t)= 2e \sum_{b,b'} {f_{b,b'} \L _{b,b'} ^{(\a)} 
\sin[(\epsilon_{b}^{\inf} -\epsilon_{b'}^{\inf})t]}.
 \label{dcurr} 
 \end{equation} 
In Eq.~(\ref{dcurr}) the summation is over all bound states of the final Hamiltonian 
$\bH ^{\inf}$ and $I_{\a} ^{(D)}$ oscillates with frequencies given by the 
differences of the bound-state eigenenergies. The quantities 
${\L}_{b,b'}$ and $f_{b,b'}$ are 
defined according to  
 \begin{equation}
{\L}_ {b,b'} ^{(\a)}=\Tr_C \left[|\psi_{bC}^{\inf}\ket \bra 
\psi_{b'C}^{\inf}|\bgS _{\a}^{A}{(\epsilon_b')} \right],
\label{lambb'}
\end{equation} 
and 
\begin{equation}
f_{b,b'}= \bra \psi'_{b }|f(\bH^{0})|\psi'_{b' } \ket \; .
\label {fbb'}
\end{equation}
The state $| \psi_{bC}^{\inf} \ket$ is the projection of the bound 
eigenstate $| \psi_{b}^{\inf} \ket$ of the biased Hamiltonian $\bH^{\inf}$ 
onto the central region. The state  $|\psi'_{b } \ket$ is related to 
$|\psi_{b }^{\inf} \ket$ by a unitary transformation:
\begin{equation}
    \left[\begin{array}{c}
	|\psi'_{bL}\ket \\ |\psi'_{bC}\ket \\ |\psi'_{bR}\ket
      \end{array}
      \right]=
    \left[
      \begin{array}{ccc}
	e^{i\D^{\inf}_{L}}\bone_{L} &0 & 0 \\
        0 &\bM_{C}&0 \\
        0 &0 &e^{i\D^{\inf}_{R}}\bone_{R}
      \end{array}
      \right] 
    \left[\begin{array}{c}
	|\psi_{bL}^{\inf}\ket \\ |\psi_{bC}^{\inf}\ket \\ |\psi_{bR}^{\inf}\ket
      \end{array}
      \right],
     \label {unittrans}
\end{equation}
with
\begin{equation}
\D^{\inf}_{\a}=\lim_{t\rightarrow \inf} \int _{0}^{t} {\rm d}t' 
(U_{\a} \left(t' \right)-U_{\a}^{\inf}),
 \label {biasmem}
\end{equation} 
$\bM_{C}$ a unitary ``memory matrix'' with the same dimension as 
the number of degrees of freedom employed to describe 
region $C$ and $\bone_{\a}$  the identity matrix 
projected onto region $\a= L,R$ . The memory matrix depends on the history of the 
time-dependent perturbation and is defined through the equation below
\be
\lim_{t\ra\inf}\mathbf{G}^{A}_{CC}(0;t)=
\bM_{C}\lim_{t\ra\inf}\bar{\mathbf{G}}^{A}_{CC}(0;t),
\label{memc}
\ee
where $\bar{\mathbf{G}}^{A}_{CC}(0;t)$ is the projection onto region 
$C$ of the advanced Green function $\bar{\mathbf{G}}^{A}(0;t)=i\exp(i\bH^{\inf}t)$.

Few remarks about the central result in Eq.~(\ref{s+dcurr}) are in order. 
First, we wish to emphasize again that {\em no steady-state current} develops 
if the biased Hamiltonian $\bH^{\inf}$ has bound eigenstates. The current 
oscillations given by Eq.~(\ref{dcurr}) are persistent, i.e., they do not decay 
in time. Second, in contrast 
to the case without bound states, the asymptotic current $I_{\a}(t)$
depends both on the initial equilibrium configuration and 
history of the applied bias and gate voltage through the coefficients 
$f_{b,b'}$ of Eq.~(\ref{fbb'}).
For sudden 
switching of the bias and gate voltage $\D^{\inf}_{\a}=0$ and 
$\bM_{C}={\bone}_{C}$ (${\bone}_{C}$ being the identity matrix 
projected onto region $C$) and 
the matrix in Eq.~(\ref{unittrans}) reduces to the identity matrix. 
On the contrary, different switching processes yield different memory 
matrices and hence different amplitudes of the current oscillations, 
see Section \ref{num} for a detailed study of the history dependence. 
Third, the NEGF formalism described in this Section 
can be combined with Time-Dependent Density Functional Theory
\cite{RungeGross:84,GrossKohn:90} (TDDFT) to include exchange and 
correlation effects in the calculated density 
and current. In this theory the steady-state assumption is consistent with 
the TDDFT equation for the total current provided the density 
of states in region $C$ is a smooth function.\cite{StefanucciAlmbladh:04}
On the contrary, the presence of bound-states in the biased Hamiltonian 
is not compatible with a steady current.\cite{Stefanucci:07}
This result opens up the possibility of having 
oscillatory solutions even for constant biases and may change 
substantially the standard steady-state picture already at the level of 
exchange-correlation functionals which are local or semi-local in time. 
On one hand, the oscillations of the effective potential in 
region $C$  give rise to new conductive channels, an effect that 
cannot be captured in any static approach. On the other hand, 
the asymptotic ($t\ra\inf$) density depends on the occupation coefficients 
$f_{b,b}$ which in turn depend on the history of the TDDFT potential. 
Thus, history-dependent effects might be observed even at the level of the 
adiabatic local density approximation.
Finally we emphasize that the above conclusions are not limited 
to TDDFT but also apply to any other single-particle theory of electrons 
such as, e.g., Hartree-Fock theory. Similarly, they also apply to a 
single-electron theory of coupled 
electronic and nuclear motion where the time evolution of the nuclei is 
treated in the Ehrenfest approximation and thus the potential acting on the 
electrons depends parametrically on the (time-dependent) nuclear 
coordinates. In this latter case the presence of a self-consistent 
oscillatory solution in a Holstein wire connected to one-dimensional 
non-interacting leads was observed in 
Ref.~\onlinecite{VerdozziStefanucciAlmbladh:06}.
\subsection{Direct propagation of the time-dependent Schr\"odinger equation}

Calculating the time-dependent current in terms of the Green function  
projected onto the central region amounts to solving either the 
Keldysh-Dyson integral equations\cite{zmjgw.2005,mwg.2006} or 
the integro-differential Kadanoff-Baym 
equations.\cite{KadanoffBaym:62,dvl.2007} In this work we use an alternative 
approach which is based on solving the  
time-dependent Schr\"odinger equation for the initially occupied 
one-particle states.\cite{KurthStefanucciAlmbladhRubioGross:05} 
An advantage of the latter approach over the former ones is that the 
wave-functions depend only on one time argument as opposed to the 
double time dependence of the Green function. This algorithm has recently 
also been used to study electron pumping by direct time propagation 
\cite{StefanucciKurthRubioGross}. 

For non-interacting electrons 
at zero temperature the total current from region $\a$ of Eq.~(\ref{cudft}) 
can alternatively be expressed as a surface integral 
\begin{equation}
I_{\a}(t)= -e\sum_{\rm occ}\int_{S_{\a}}\dr\s\,\hat{{\bf n}}\cdot 
{\rm Im}\left[\q_{n}^{\ast}(\br,t)\grad\q_{n}(\br,t)\right],
\label{currgen}
\end{equation}
where $\hat{{\bf n}}$ is the unit vector perpendicular to the surface 
element ${\rm d}\s$, the surface $S_{\a}$ is perpendicular to the longitudinal 
geometry of our system and $\q_{n}(\br,0)$ are the eigenstates 
of $\bH(t<0)$. The electrode-junction-electrode system is infinitely 
extended and non-periodic. In practice, of course, we can only deal 
with finite systems and therefore we only propagate the initial 
wavefuction projected onto the central region $C$. The presence of the leads 
is taken into account by applying the correct boundary conditions. 
It is worth to note that even for interacting electrons one can use 
Eq.~(\ref{currgen}) to compute the current through the junction if the 
single-particle orbitals $\q_{n}(\br,t)$ are the Kohn-Sham orbitals of 
time-dependent density functional theory. 

For a description of the algorithm proposed in 
Ref.~\onlinecite{KurthStefanucciAlmbladhRubioGross:05}, it is convenient to 
write $\bH_{\a\a}(t)$, with $\a=L,R$ , as the sum of a term 
$\bH^{0}_{\a\a}=\bH_{\a\a}(0)$ which is constant in time and another term 
$\bU_{\a}(t)$ which may be explicitly time-dependent, 
$\bH_{\a\a}(t)=\bH^{0}_{\a\a}+\bU_{\a}(t)$. 
In configuration space $\bU_{\a}(t)$ 
is diagonal at any time $t$ since the potential is local in space.
Furthermore, the diagonal elements $U_{\a}(\br,t)$ are spatially 
constant for metallic electrodes. Thus, 
$\bU_{\a}(t)=U_{\a}(t){\bone}_{\a}$ and $U_{L}(t)-U_{R}(t)$ is the 
total potential drop across the junction. 
The total Hamiltonian is $\bH(t)=\tilde{\bH}(t)+\bU(t)$ with 
\begin{equation}
\tilde{\bH}(t)=\left[
\begin{array}{ccc}
\bH^{0}_{LL} & \bH_{LC} & 0 \\
\bH_{CL} & \bH_{CC}(t) & \bH_{CR} \\
0 & \bH_{RC} & \bH^{0}_{RR}
\end{array}
\right]
\nonumber 
\end{equation}
and
\begin{equation}
\bU(t)=\left[
\begin{array}{ccc}
U_{L}(t){\bone}_{L} & 0 & 0 \\
0 & 0 & 0 \\
0 & 0 & U_{R}(t){\bone}_{R}
\end{array}
\right].
\end{equation}
In this way, the only term  
in $\tilde{\bH}(t)$ that depends on $t$ is $\bH_{CC}(t)$. 
For any given initial one-particle state $|\q(0)\ket=|\q^{(0)}\ket$ we calculate 
$|\q(t_{m}=m\D t)\ket=|\q^{(m)}\ket$ by employing a generalized form of 
the Cayley method (atomic units are used throughout)
\begin{equation}
\left({\bf 1}+i\d \tilde{\bH}^{(m)}\right)
\frac{{\bf 1}+i\frac{\d}{2}\bU^{(m)}}
{{\bf 1}-i\frac{\d}{2}\bU^{(m)}}|\q^{(m+1)}\ket=
\nonumber
\end{equation}
\begin{equation}
\left({\bf 1}-i\d \tilde{\bH}^{(m)}\right)
\frac{{\bf 1}-i\frac{\d}{2}\bU^{(m)}}
{{\bf 1}+i\frac{\d}{2}\bU^{(m)}}|\q^{(m)}\ket,
\label{prop}
\end{equation}
with 
$\tilde{\bH}^{(m)}=\frac{1}{2}[\tilde{\bH}(t_{m+1})+\tilde{\bH}(t_{m})]$, 
$\bU^{(m)}=\frac{1}{2}[\bU(t_{m+1})+\bU(t_{m})]$ and $\d=\D t/2$. The above   
propagation scheme is unitary (norm conserving) and accurate to   
second-order in $\d$. From Eq.~(\ref{prop}) we can extract an equation   
for the time-evolved state in region $C$. After some algebra, one ends up with 
an equation which gives the wave function in region $C$ 
at time step $m+1$ in terms of the wave function in region $C$ at the previous time 
step and two additional terms (source and memory term): 
\begin{equation}
|\q_{C}^{(m+1)}\ket=
\frac{{\bf 1}_{C}-i\d \bH_{\rm eff}^{(m)}}{{\bf 1}_{C}+i\d \bH_{\rm eff}^{(m)}}
|\q_{C}^{(m)}\ket
+|S^{(m)}\ket-|M^{(m)}\ket.
\end{equation}
The effective Hamiltonian $\bH_{\rm eff}^{(m)}$ of region $C$ is 
defined according to
$\bH_{\rm eff}^{(m)}=\bH_{CC}^{(m)}-
i\d\bH_{CL}(1+i\d\bH^{0}_{LL})^{-1}\bH_{LC}
-i\d\bH_{CR}(1+i\d\bH^{0}_{RR})^{-1}\bH_{RC}$, where
$\bH_{CC}^{(m)}=\frac{1}{2}[\bH_{CC}(t_{m+1})+\bH_{CC}(t_{m})]$.
The source term $|S^{(m)}\ket$ depends on the initial wavefunction 
in region $\a=L,R$ and reads 
\beq
  |S^{(m)}\ket=-\frac{2i\d}{1+i\d \bH_{\rm eff}^{(m)}} 
  \sum_{\a=L,R}\frac{\L_{\a}^{(m,0)}}{u_{\a}^{(m)}}\bH_{C\a}
  \nonumber \\ \times
  \frac{(1-i\d\bH_{\a\a})^{m}}{(1+i\d\bH_{\a\a})^{m+1}}|\q_{\a}^{(0)}\ket \;,
  \label{source}
\eeq
with
\begin{equation}
u_{\a}^{(m)}=\frac{1-i\frac{\d}{2}U_{\a}^{(m)}}{1+i\frac{\d}{2}U_{\a}^{(m)}}
\quad {\rm and}\quad
\L_{\a}^{(m,k)}=\prod_{j=k}^{m}[u_{\a}^{(j)}]^{2}.
\end{equation}
The memory term $|M^{(m)}\ket$ is responsible for the hopping in and 
out of region $C$. It depends on the wavefunction in the device region 
at previous time steps and reads
\begin{eqnarray}
\lefteqn{
  M^{(m)}= -\frac{\d^{2}}{1+i\d \bH_{\rm eff}^{(m)}}\sum_{\a=L,R}
\sum_{k=0}^{m-1} \frac{\L_{\a}^{(m,k)}}{u_{\a}^{(m)}u_{\a}^{(k)}}} \nonumber \\
&&  \left[\bQ_{\a}^{(m-k)}+\bQ_{\a}^{(m-k-1)}\right]
 \left(|\q_{C}^{(k+1)}\ket+|\q_{C}^{(\k)}\ket \right) \; ,
  \label{memory}
\end{eqnarray}
with 
$\bQ_{\a}^{(m)}=\bH_{C\a}[(1-i\d\bH_{\a\a})^{m}/(1+i\d\bH_{\a\a})^{m+1}]\bH_{\a C}$.
For more details on the implementation of the algorithm the reader is referred 
to Ref.~\onlinecite{KurthStefanucciAlmbladhRubioGross:05}. 

\section{Numerical results}  
\label{num}

In this Section we present the results of our numerical simulations for simple 
one-dimensional model systems which support two bound states in the long-time 
limit. Of particular interest will be the dynamical part of the current and 
the dependence of the amplitude of the bound-state oscillations on the history of the time-dependent potential 
and on the initial state. We also identify single-particle transitions other than 
between the bound states which are relevant to understand the shape of the 
transient current.

The time-dependent, one-dimensional Hamiltonian is given by 
\begin{equation}  
    H(x,t)=-\frac{1}{2} \frac{{\rm d}^{2}}{{\rm d}x^2}+U_0(x)+U(x,t) =: H^{0}(x) + U(x,t)\;.  
\end{equation}
For times $t\leq 0$ the Hamiltonian is $H^{0}(x)$ and the system is in its 
ground state. At $t=0$ the system is driven out 
of equilibrium by the time-dependent potential $U(x,t)$. We choose the 
time-dependent perturbation in such a way that for ${t\rightarrow \inf}$ the 
Hamiltonian globally converges to an asymptotic Hamiltonian, which we 
denote with $H^{\inf}(x)$. 

The time-dependent perturbation $U(x,t)$ can be written as a piece-wise 
function of the space  variable $x$. Let $U_{\a}(t)$ be the applied 
bias in region $\a=L,R$ and $V_g(x,t)$ the gate voltage applied to 
region $C$. The latter may depend on both position $x$ and time $t$.
Then
 \begin{equation}
       U(x,t) = \left \{ \begin{array} {lc}
       U_{L}(t) &       -\inf < x < x_{L}\\
       V_g(x,t)  &       x_L < x < x_{R}\\
       U_{R}(t)   &    x_{R}< x < \inf
         \end{array}\right. ,
 \end{equation}
with $x_{L}$ and $x_{R}$ the positions of the left and right 
interfaces respectively.
In our numerical implementation we discretize $H$ on a equidistant grid and 
use a simple three-point discretization for the kinetic energy. 
In all systems studied below the simulations have been performed by 
considering a propagation window which extends from $x_L=-1.2$ a.u. 
to $x_R=1.2$ a.u. and a lattice spacing $\D x=0.012$ a.u..
The occupied part of the continuous spectrum ranges
from $k=0$ to $k_{\rm F}=\sqrt{2\ve_{\rm F}}$ and it is discretized with 
200 $k$-points. All occupied states are propagated from 
$t=0$ to $t=1400$ a.u. using a time step $2\d=0.05$ a.u.. In all the numerical 
examples studied below the final Hamiltonian supports two bound states and the 
resulting current in the long-time limit then is
\begin{equation}
I(t) = I^{(S)} + I_{\rm osc}(x) \sin ( \w_{0} t)
\end{equation}
and, on top of the steady current $I^{(S)}$, has an oscillating part with 
only one frequency $\w_{0}$ given by the eigenenergy difference of the two 
bound states. It is also worth mentioning that the amplitude $I_{\rm osc}$ 
of this current oscillation depends on the position (see Eq.~(\ref{lambb'})) 
while the steady-state current is position-independent.

\subsection{Bound state oscillations and transients}

As a first example, we study a system with an initial potential 
$U_{0}(x)=0$. Initially, the system is in the ground state with Fermi 
energy $\ve_{\rm F}=0.1$ a.u.. All wavefunctions of the ground-state 
Slater determinant are extended one-particle states with energy 
between 0 and $\ve_{\rm F}$. 
At $t=0$, the system is suddenly driven out of equilibrium by 
switching on a potential $U(x,t)$ which consists of a constant bias in the left lead, 
$U_{L}=0.1$ a.u., and a constant gate voltage in the central region, $V_g=- 1.4$ a.u.. 
\begin{figure}[tb]  
\includegraphics*[width=.45\textwidth]{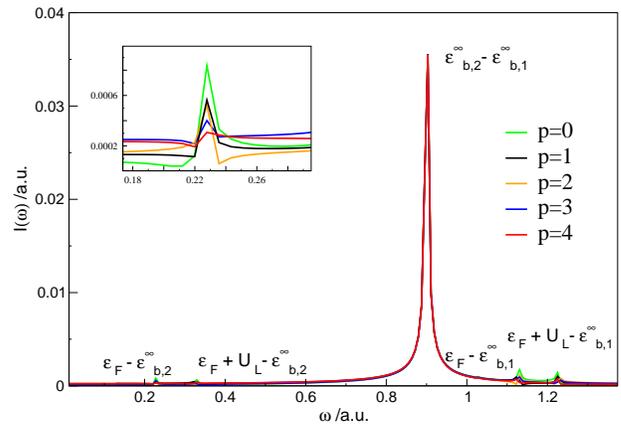}  
\caption{Modulus of the discrete Fourier transform of the current for $V_{g}=-1.4$ a.u. and a 
constant bias in the left lead
$U_{L}=0.1$ a.u.. The inset shows a magnification of   
the region with bound-continuum transitions from the bound state with higher energy to the Fermi energy. Different curves   
correspond to different time intervals.}  
\label{transientdc}  
\end{figure}  
The biased Hamiltonian has two bound eigenstates with energies 
$\ve_{b,1}^{\inf}=-1.032$ a.u. and $\ve_{b,2}^{\inf}=-0.133$ a.u.. 
From the discussion of the previous Section we expect that a steady 
state cannot develop and that the time-dependent current exhibits an 
oscillatory behavior with frequency 
$\w_{0}=\ve_{b,2}^{\inf}-\ve_{b,1}^{\inf}$. This is indeed confirmed by our 
numerical simulations, as one can see in Fig.~\ref{transientdc} where 
we plot the modulus of the discrete Fourier transform of the time-dependent 
current. The latter quantity is defined according to
\begin{equation}  
    I(\w_{k})=\frac{2\d}{\pi\sqrt{2N_{0}}}\sum_{n=n_{p}}^{n_{p}+N_{0}}I(2n\d)  
    e^{-i\w_{k} n\d },  
    \quad \w_{k}=\frac{2\p k}{N_{0}\d} . 
    \label{dftc}  
\end{equation}
We have computed ${I}(\w_{k})$ for different values of 
$n_{p}=(4+2p)\cdot 10^{3}$, $p=0,1,2,3,4$, 
and $N_{0}=16\cdot 10^{3}$. 
Different values of $p$ correspond to different time intervals 
$t\in(t_{p},t_{p}+T_{0})$ with 
$t_{p}=(2+p)\times 100$ a.u. but with the {\em same} duration $T_{0}=800$ a.u.. 
The coefficient in Eq.~(\ref{dftc}) is defined such that the height of the peak $I(\w)$ at $\w$ is equal to the amplitude 
of the oscillations with frequency $\w$.
Besides the zero-frequency peak (not shown) due to the non vanishing dc 
current, ${I}(\w)$ shows a dominant peak at the frequency 
$\w_0 = \ve_{b,2}^{\inf}-\ve_{b,1}^{\inf}$ of the transition between the 
two bound states. As expected, the height of this peak remains unchanged 
as $p$ varies from 0 to 4, i.e., the current 
oscillation associated with this transition remains undamped. 
We emphasize that they are an intrinsic 
property of the biased system.

Closer examination of Fig.~\ref{transientdc} reveals four extra peaks which 
are related to different internal transitions. The first and the last pairs 
of peaks occur at frequencies which correspond to transitions between the 
bound states and the lower edge of the unoccupied part of the continuous spectrum in 
the left and right lead of the biased system, 
$\ve_{b,i}^{\inf}\rightarrow \ve_{\rm F}$, and   
$\ve_{b,i}^{\inf}\rightarrow \ve_{\rm F}+U_{L}$, with $i=1,2$. These   
sharp structures (mathematically stemming from the discontinuity of the   
zero-temperature Fermi distribution function)   
give rise to long-lived oscillations of the total current and density.  
These oscillatory transients die off very slowly, the height of the peaks 
decreases with increasing $t_{p}$ empirically as $1/t_p$ (power-law 
behavior).
\begin{figure}[tb]  
\includegraphics*[width=.45\textwidth]{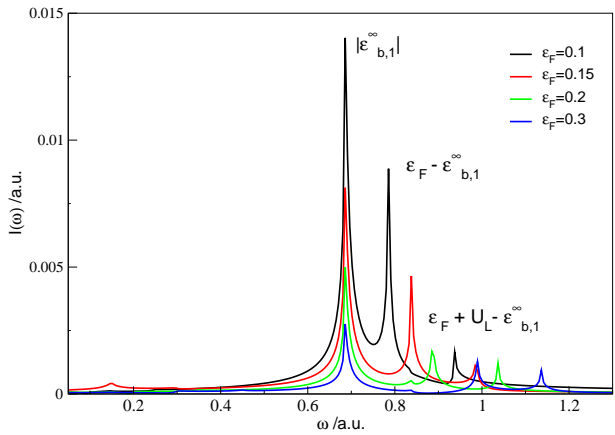}  
\caption{Modulus of the discrete Fourier transform of the current of a translationally invariant initial Hamiltonian which is perturbed at $t=0$ by a sudden bias in the left lead $U_{L}=0.15$ a.u. and the system evolves toward a steady state. Then, at $T=150$ a.u. a gate voltage 
$V_{g}(x)=-v_{g}=-1.02$ a.u. is suddenly turned on. The first peak appears at the $\w=0.686$ a.u. which is the modulus of the energy of the bound eigenstate 
of the final Hamiltonian ($H(x,t>T)$ has one bound eigenstate). Different 
curves correspond to different Fermi energies.}  
\label{transient}  
\end{figure} 
In Fig.~\ref{transientdc}, as well as in all following examples, we  
report results for the current calculated in the center of the device 
region. However it is 
worth to mention that the amplitude of the current oscillations decays 
exponentially in the leads as $e^{-(k^{\a}_{b,1}+k^{\a}_{b,2})|x-x_{\a}| }~$  where 
$ k^{\a}_{b,i} = \sqrt {2 (|\ve_{b,i}^{\inf}|+U_{\a})}$ with $i=1,2$, $\a =L,R$ 
and $x$ is a point in lead $\a$. Consequently, the dynamical part of the 
current vanishes deep inside the leads (away from where the 
bound states are localized).

In the second example, we consider a system described by the 
translationally invariant Hamiltonian 
$H(x,t<0)=-\frac{1}{2}\frac{\dr^{2}}{\dr x^{2}}$. At $t=0$ we suddenly switch
on a constant bias in the left lead $U_L=0.15$ a.u. and propagate until 
$T=150$ a.u. when a steady state is reached. At $t=T$ a gate voltage 
$V_{g}(x)=-v_{g}=-1.02$ a.u. is suddenly turned on and the Hamiltonian 
$H(x,t>T)$ has one bound eigenstate at energy $\ve^{\inf}_{b}=-0.686$ a.u. . 
The depth $v_{g}$ is chosen in such a way that if one slightly 
increases $v_g$ a second bound eigenstate appears. Since the
system has only one bound state, the oscillations die out
slowly as $1/(t-T)$ and eventually another steady state develops. 
In order to understand the transient oscillations we have studied 
the Fourier transform of the current as shown in Fig.~\ref{transient}. There  
the first peak appears at the frequency of $\w= |\ve_{b,1}^{\inf}|$  which is 
a transition between the bound level and the bottom of
the continuum. As such, the position of this peak remains unchanged for 
different Fermi energies. Besides this transition one observes other 
peaks whose positions shift as the Fermi energy is changed. 
They correspond to transitions from the bound level to the top of the 
left and right continua and, as for the first transition, they decay as 
$1/(t-T)$. 

\subsection{Dependence of the current oscillations on the initial conditions}
\label{depin}  

\begin{figure}[htbp]  
\includegraphics*[width=.45\textwidth]{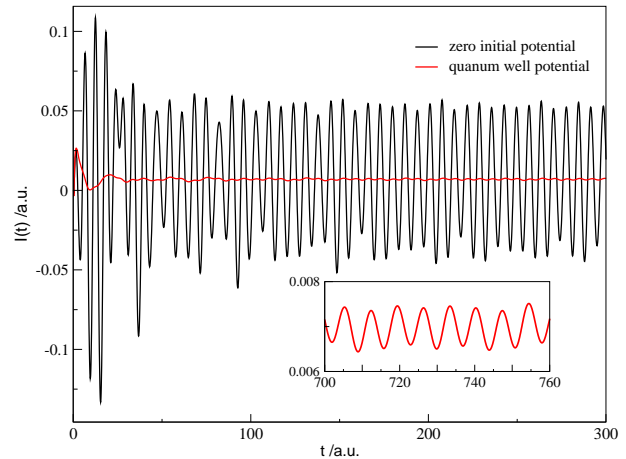}
\caption{Comparison of the time-dependent current for systems with and 
without bound states at negative times.  The inset 
shows a magnification of the time-dependent current of the system with 
two initial bound states. Since both systems have the same final Hamiltonian, 
the frequencies of the current oscillations are the same while the amplitude 
of the oscillations for the quantum well (with two bound state initially) is 
smaller by almost two orders of magnitude than for the system without initial bound states.}
\label{initcon}    
\end{figure}

The dynamical part of the current depends on the initial Hamiltonian  
$H^{0}(x)$ through the amplitudes $f_{b,b'}$ of Eq.~(\ref{fbb'}). In 
the first example of the previous Section the Hamiltonian at 
negative times, $H^{0}(x)$, had no bound eigenstates. At positive 
times a gate voltage and a bias in the left lead were suddenly switched on
and the Hamiltonian at positive times is equal to $H^{\inf}(x)$ and has two bound 
eigenstates. 
We now consider a system with two bound eigenstates for $t\leq 0$ and 
exposed to a dc bias for $t>0$. Specifically, we start with a static 
potential describing a quantum well of depth $U_{0}(x)=-1.4$ a.u. 
for $|x|<1.2$ a.u.. The ground state of the system is the Slater 
determinant of all the extended eigenstates with energy up to 
$\ve_{\rm F}=0.1$ a.u. and of the two bound eigenstates at energies 
$\ve^{0}_{b,1}=-1.035$ a.u. and $\ve^{0}_{b,2}=-0.156$ a.u..
At $t=0$ a dc bias $U_R = 0.1$ a.u. is suddenly switched on in the 
left lead and the Hamiltonian $H(x,t>0)=H^{\inf}(x)$ is equal to the 
final Hamiltonian studied in the previous Section. The resulting 
time-dependent current for these two systems are shown in Fig.~\ref{initcon}. 

As a consequence of the fact that $H^{\inf}(x)$ is the same in both 
systems the time-dependent currents should oscillate with 
the same frequency, a result which is confirmed by our numerical calculation. 
The amplitude of this oscillation, however, depends on the 
initial equilibrium configuration as well as on how $H(x,t)$ approaches the 
asymptotic Hamiltonian $H^{\inf}(x)$.
As one can see from Fig.~\ref{initcon}, the amplitude is much 
larger in the system with no initial bound states. This difference can 
be explained qualitatively by looking at Eq.~(\ref{fbb'}). In both 
systems the time-dependent perturbation is switched on suddenly.
Therefore, the transformation matrix of Eq.~(\ref{unittrans}) 
becomes the unit matrix and Eq.~(\ref {fbb'}) reduces to   
\begin{equation}
f_{b,b'}=\bra\psi^{\inf} _{b}|f(\bH^{0})|\psi^{\inf}_{b'}\ket.
\end{equation}
When the perturbation is small like in the case of the system with 
two initial bound states ($\bH^{0} \approx \bH^\inf$), the eigenfunctions
$|\psi_{b }^{\inf}\ket$ of $\bH^\inf$ are approximate eigenfunctions 
of $\bH^0$ as well. 
Therefore $f(\bH^{0})|\psi_{b }^{\inf}\ket \approx f(\ve_{b})
|\psi_{b }^{\inf}\ket$ and $f_{b,b'} \approx f(\ve_{b})  
\d _{b,b'}$ which leads to a vanishing dynamical part of the current since there only the off-diagonal elements contribute.
By contrast, if the applied potential $U(x,t)$ is large, the overlap 
$\bra\psi^{\inf} _{b}|f(\bH ^{0})|\psi^{\inf}_{b'}\ket$ can be quite substantial 
and the resulting amplitude of the current oscillation is large.

\subsection{Dependence of the current oscillations on the history of the bias}
\label{depbias}

The amplitude of the bound state oscillations depends, through 
the transformation matrix in Eq.~(\ref{unittrans}), on the history of the 
time-dependent potential which perturbs the initial state. In this 
Section we investigate for the first time how such amplitudes depend on 
the switching process (history-dependence effects). 

We take the flat potential $U_0(x)=0$ as initial potential and the Fermi 
energy $\e _{\rm F}=0.2$ a.u.. At $t=0$ a gate voltage $V_g (x)=-1.3$ 
a.u. abruptly lowers the potential in 
the center. In addition, 
a time-dependent bias is applied to the left lead as $U_L(t)=U_L 
\sin ^2 (\w_{b} t)$ for $t \leq t_{b} =\frac { \p}{2\w_{b}}$ and 
$U_L(t)=U_L$ for $t >\frac { \p}{2\w_{b}}$, where 
$U_L=0.1$ a.u.. 

The final biased Hamiltonian has two bound states with 
energies $\e _{b,1}^{\inf} = -0.933$ a.u. and $\e _{b,2}^{\inf}=-0.063$ a.u. which 
again leads to undamped oscillations in the current.
\begin{figure}[t]  
\includegraphics*[width=.45\textwidth]{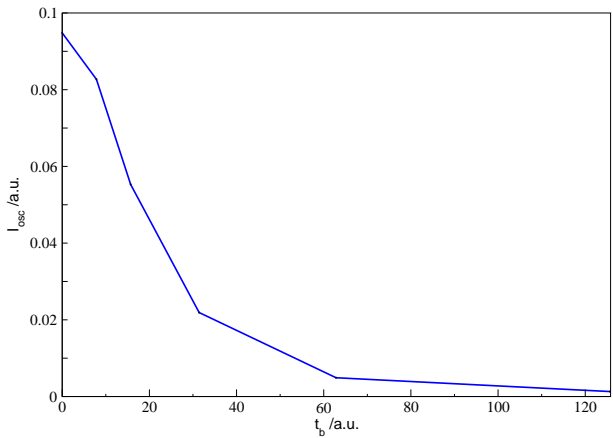}
\caption{The amplitude of the current oscillation as function of the switching 
time of the bias. The bias in the left lead is switched according to 
$U_L(t)=U_L \sin ^2 (\w_{b} t)$ for $t \leq t_{b} =\frac {\p}{2\w_{b}}$ 
and $U_L(t)=U_L=0.1$ a.u. for later times. The frequency of the 
current oscillation
$\w_0=\e _{b,2}^{\inf}-\e _{b,1}^{\inf}$ is given by the difference of bound 
state energies in the final system which have the values $\e _{b,1}^{\inf} 
= -0.933$ a.u. 
and $\e _{b,2}^{\inf}=-0.063$ a.u., respectively. The Fermi energy is $\e _{\rm F}=0.2$ a.u. and the gate potential is $V_g=-1.3$ a.u..}
\label{mxbias}  
\end{figure} 

Choosing $t_{b} $ in such a way that $\D^{\inf}_{L}$ equals $2\p$, 
$4\p$,\ldots the upper block of the unitary matrix in Eq.~(\ref{unittrans})
 become the identity matrix in region $L$.
This suggests that the amplitude of the current oscillations 
might exhibit a non-monotonic behavior as a function of the switching time. 
Our numerical results demonstrate that this is not the case. 
Fig.~\ref{mxbias} shows that the amplitude decreases monotonically as a function of 
$t_{b}$, a trend which is expected in the region of long switching times (adiabatic switching).
Such behavior, however, does not contradict the analytic results of Section 
\ref{thp}. In fact, the memory matrix in the central region $\bM_{C}$ also depends on the way 
the bias is switched on through the time-dependent embedding 
self-energy needed to calculate $\bG^{A}_{CC}(0;t)$, see Eq.~(\ref{memc}), and, in general, $\bM_{C}\neq {\bone}_{C}$ when 
$\D_{L}^{\inf}=2\p,\; 4\p,\ldots$

\subsection{Dependence of the current oscillations on the history of the gate voltage}  
\label{depgate} 

Finally we present some results to illustrate the dependence of the current 
oscillations on the switching process of the gate voltage.
\begin{figure}[htbp] 
\includegraphics*[width=.5\textwidth]{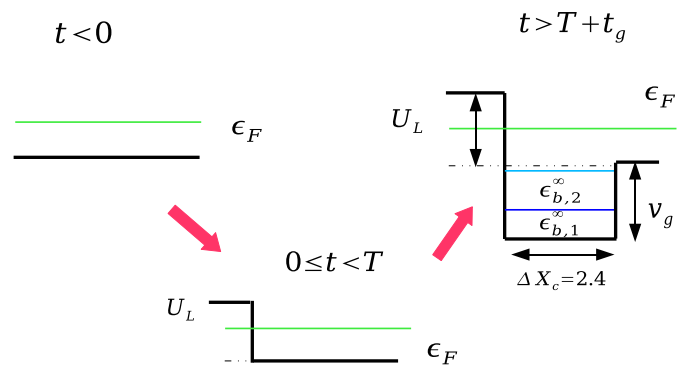}
\caption{Schematic sketch of the time evolution of the Hamiltonian. Starting 
from an initially constant potential (left), at $t=0$ a bias is 
suddenly applied 
to the left lead and the system evolves toward a steady state (center). 
Then, between times $T$ and $T+t_g$, a time-dependent gate 
voltage $V_{g}(x,t) = -\frac {v_{g}}{t_{g}}(t-T)$ is 
switched on in region $C$. For times $t>T+t_g$ (right) the Hamiltonian
remains constant in time.}
\label{gatehis}      
\end{figure}
Again we start with the constant potential $U_0(x)=0$ at equilibrium. At 
$t=0$ a bias is ramped up abruptly in the left lead and the time-dependent 
current goes through some transient which lasts for a few tens of atomic 
units. We wait long enough, a time $T=150$ a.u., for a steady-state to develop.
After this time all dependence on the history of the applied bias is washed out.\begin{figure}[htbp]
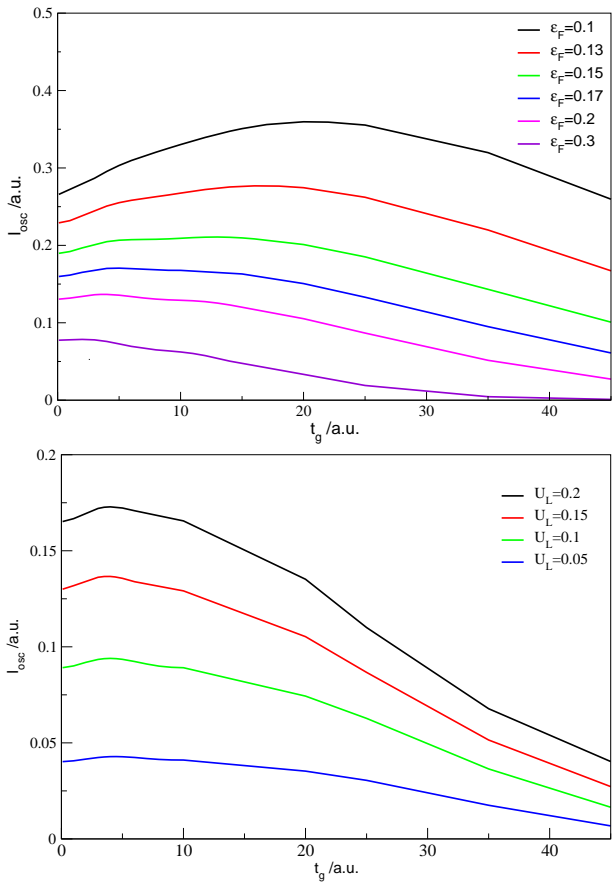
 
\includegraphics[angle=0,width=0.45\textwidth,clip]{fig6up.eps} 
\includegraphics[angle=0,width=0.45\textwidth,clip]{fig6dn.eps}
\caption{The amplitude of the current oscillations as function of the 
switching time $t_{g}$ for $v_g=1.3$ a.u.. Upper panel: for 
fixed bias $U_{L}=0.15$ a.u. and different Fermi energies. Lower panel: for 
fixed Fermi energy $\ve_{\rm F}=0.2$ a.u. and different values of the bias. 
All curves reach a maximum whose position remains almost unchanged.}  
\label{uf-fix}     
\end{figure}
 
At $t=T$ a time dependent gate voltage $V_{g}(x,t) = 
-\frac {v_{g}}{t_{g}}(t-T)$ is applied to region $C$. The gate 
voltage decreases linearly until $t=T+t_{g}$ and remains constant and 
equal to $-v_{g}$ for all later times. In Fig. \ref{gatehis} we 
provide a schematic sketch of the overall time-dependent perturbation.

The time $t_{g}$ is the 
switching time. The final Hamiltonian  $H^{\inf}(x)=H(x,t>T+t_g)$ 
has two bound eigenfunctions and the steady-state cannot develop.

In Fig.~\ref {uf-fix} the amplitude of the oscillation versus the switching 
time $t_g$ is shown for a final depth of the gate $v_g=1.3$ 
a.u.. In the upper panel, the bias in the left lead is fixed to $ U_{L}=0.15$ 
a.u. and the Fermi energy is varied from $\ve_{\rm F}=0.1$ a.u. to $0.3$ 
a.u..
We see that the amplitude reaches a maximum value for a certain 
switching time. It is also worth noting that the amplitudes 
are generally smaller for larger Fermi energies, a behavior which   
can be explained as follows: let $|\f_{n}\ket$ be an eigenstate of 
$\bH^{0}$ with eigenenergy $\ve_{n}$. Then 
\be
f_{b,b'}=\sum_{\ve_{n}<\ve_{\rm F}}
\bra\psi'_{b}|\f_{n}\ket\bra\f_{n}|\psi'_{b'}\ket.
\ee
As the Fermi energy increase the sum over $\ve_{n}$ approaches the 
sum over a complete set of eigenstates and hence $f_{b,b'}$ 
approaches the value $\bra\q'_{b}|\q'_{b'}\ket$. This latter quantity 
vanishes since the states $|\q'_{b}\ket$ are related to the 
orthogonal states $|\q^{\inf}_{b}\ket$ by a unitary transformation 
and hence remain orthogonal. 
The lower panel of Fig.~\ref {uf-fix} shows the amplitude versus 
the switching time of the gate voltage for a fixed Fermi energy 
$\ve_{\rm F}= 0.2$ a.u. and for different values of the applied bias. 
The striking feature of this plot is that the position of the maximum 
remains almost unchanged as function of the bias $U_L$. 
         
\begin{figure}[tb] 
\includegraphics[angle=0,width=0.45\textwidth,clip]{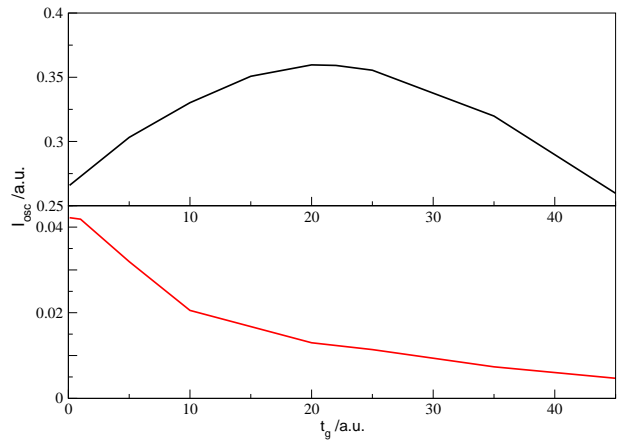} 
\caption{The amplitude of the current oscillation as function of the switching 
time of the gate. The red (black) curve refers to the initial ground 
state with (without) a bound state. The numerical parameters are  
$\ve_{\rm F} =0.1$ a.u., $U_L=0.15$ a.u.}
\label{ant}       
\end{figure}

As a final example, in Fig.~\ref{ant} we compare the amplitude of the 
oscillations as function of the switching time $t_{g}$ for two different 
initial states with the same Fermi energy $\ve_{\rm F}=0.1$ a.u. 
In one case we start, as before, with the constant potential
$U_{0}(x)=0$, and hence $H^{0}(x)$ does not have bound eigenstates.
In the other case we start with a quantum well of depth $U_{0}=- 0.5$ a.u. for
$|x|\leq 1.2$ a.u.. The Hamiltonian $H^{0}(x)$ in this latter case 
has one bound eigenstate. A bias $U_L=0.15$ a.u. in the left lead 
is suddenly switched on in both systems and after a time $T=150$ a.u. a 
steady state is attained. For $T<t<T+t_{g}$ a gate voltage $V_{g}(x,t)$ is 
gradually switched on as before, and for $t>T+t_{g}$ the gate voltage
remains constant and equal to $v_{g}=-1.3$ a.u. in the first case 
and $-0.8$ a.u. in the second case. Hence, both systems have the same 
asymptotic Hamiltonian $H^{\inf}(x)$. The remarkable difference between the value of the amplitudes in these cases can be explained in the same way as in Section \ref{depin}.  

Interestingly, in the case where the system initially has one bound state, 
the amplitude has a maximum for sudden switching of the gate, i.e., 
$t_g=0$ a.u., while in the case with no initial bound states the maximum 
appears at a finite value of $t_g$. 

Similarly, we have found a maximum for small  
$t_g$ for the following situation: we start with an initial state 
without bound states. At $t=0$ a.u. we suddenly apply a bias in the left lead 
and wait until a steady state is achieved. Then we switch on a gate 
in such a way that {\em one} bound state is created and wait until the 
associated bound-continuum transitions have decayed before we add another 
bound state to the gate with a switching time $t_g$. 

The fact that in this case the largest amplitude for the current oscillations 
is found for switching time $t_g$ close to zero 
strongly suggests that the 
position of the maximum in the oscillation amplitude as function of $t_g$ is 
related to a transient effect. This is also supported by the following 
observation (see Fig.\ref{uf-fix}): the switching time $t_g$ for which the 
current oscillations are largest depends on the Fermi energy (for fixed bias) 
since the transitions from the bound states to the top of the Fermi sea 
obviously 
depend on $\ve_{\rm F}$. At the same time, the position of this maximum is 
almost independent of the bias (for fixed Fermi energy) since the bias 
only leads to a slight energy shift for the bound states.

\section{Conclusions}
\label{concl}

In the theory of electron transport one usually assumes that the application 
of a dc bias to an electronic system attached to two macroscopic electrodes 
always leads to the evolution of a steady-state current. Recent theoretical 
work states \cite{Stefanucci:07} that the presence of bound states leads to 
qualitatively new features (current oscillations and memory effects) in the 
dynamics of electron transport in the long-time limit. These as well 
as transient features are investigated here in detail by numerical 
simulations. In the Fourier transform of the calculated time-dependent 
current one not only finds the predicted transitions between the bound 
states in the long-time limit, but, moreover, transitions (in the transient regime) 
between the bound states and the continuum of the leads can also be clearly 
identified. We have shown that the amplitude of the persistent current 
oscillations depends both on the initial state and on the history of system. 
Since current and density are related via the continuity equation, also the 
time-dependent density in the long-time limit will therefore be 
history-dependent. Interestingly, these memory effects show up not only in 
the dynamical part but also in the {\em time-independent} contribution of 
the bound states to the density \cite{KhosraviStefanucciKurthGross}. 

Our results indicate that in transport calculations special care has to be 
taken if bound states are present in the biased system. A warning flag has 
already to be raised at the assumption of the evolution to a steady state 
which is not true in general. Of course, the theoretical analysis predicts 
the existence of oscillations in the current but makes no statement on their 
relative importance as compared to the steady-state contribution. Our 
results show, however, that the amplitude of the oscillations locally may 
very well be comparable or even larger than the steady-state current and 
therefore cannot be neglected. We would also like to point out that the 
existence of bound states in biased transport systems may not be an exotic 
feature in an experimental situation. For single molecules attached to 
metallic leads it is quite conceivable that some of the molecular orbitals 
energetically fall into an energy window which corresponds to an energy gap 
of the leads and those orbitals therefore cannot hybridize with any lead 
states and remain fully localized. In the case of transport experiments on 
quantum dots one could artificially create bound states by applying a strong 
attractive gate potential. 

Although our numerical simulations were performed for non-interacting 
electrons, the conclusions about the dynamical current oscillations apply to 
any effective single-electron theory. In particular they also apply to the TD
Kohn-Sham equations which are in principle able to reproduce the time-dependent density\cite{RungeGross:84} (and 
the longitudinal current via the continuity equation) of an interacting system if the exact 
exchange-correlation functional is used. Intuitively, one might expect that 
electron-electron scattering leads to a 
damping of the oscillations in the long-time limit. However, the assumption 
of a time-independent density producing a static Kohn-Sham potential for large times leads to a contradiction 
if this potential supports bound states since the density and therefore 
also the Kohn-Sham potential should then become time-dependent again.

\section*{Acknowledgements}

We gratefully acknowledge useful discussions with Ali Abedi. This work was supported by the Deutsche 
Forschungsgemeinschaft, DFG programme SFB658, and the EU Network of Excellence 
NANOQUANTA (NMP4-CT-2004-500198).

\end{document}